%% file: Main.tex
\documentclass[journal]{IEEEtran}
\usepackage{caption}
\usepackage{ifpdf}
\usepackage[pdftex]{graphicx}
\usepackage{cite}
\usepackage[ruled,vlined]{algorithm2e}
\usepackage{amsmath,amssymb,amsthm}
\usepackage{hyperref}
\usepackage{mathtools}
\usepackage{algorithmic}
\usepackage{array}
\usepackage{booktabs,longtable}
\usepackage{array}
\usepackage{tikz}
\usepackage{lettrine}
\usepackage{tabularx}
\usepackage[caption=false,font=footnotesize]{subfig}
\usepackage{stfloats}
\usepackage{url}
\usepackage{svg}
\usepackage{esvect}
\usepackage{subcaption}
\usepackage{booktabs}
\usepackage{multirow}
\usepackage{float}
\usepackage{listings}
\usepackage{todonotes}
\usepackage{soul}   
\usepackage{xcolor}
\usepackage{amsthm}
\usepackage{varwidth}
\usepackage{placeins}
\usepackage{siunitx}
\usepackage{parskip}
\newtheorem{prob}{Problem}

\begin{document}
\title{Spatial Optimization of Interconnected Systems \\ in Non-Convex Design Spaces}

\author{S.~Westerhof, T.~Hofman 
\thanks{S. Westerhof and T. Hofman (e-mail: t.hofman@tue.nl) are with the Eindhoven University of Technology (TU/e), Dept. of Mechanical Engineering, \href{https://www.tue.nl/en/research/research-groups/control-systems-technology}{Control Systems Technology} section, \href{https://www.tue.nl/en/research/research-groups/group-hofman}{Engineering Systems Design} group, P.O.Box 513, 5600 MB Eindhoven, The Netherlands.}}

\maketitle
\begin{abstract}
This paper presents a spatial optimization methodology that extends the Spatial Packaging of Interconnected Systems with Physical Interaction (SPI2) framework to support arbitrary, non-convex design boundaries. We introduce a smooth, differentiable inside–outside evaluation for components represented using the Maximal Disjoint Ball Decomposition (MDBD) method. The framework also incorporates center-of-gravity and moment-of-inertia calculations directly into the optimization, and provides an end-to-end computer-aided design (CAD) workflow for importing components and reconstructing the optimized assembly. The method is demonstrated on a fictional aircraft auxiliary unit. Results show that the optimizer can place multiple interconnected components within a custom geometry while simultaneously handling routing and physics-based objectives. The approach maintains geometric feasibility within numerical tolerance and illustrates the potential of MDBD-based SPI2 methods for practical engineering design applications.
%

\end{abstract}


\section{Introduction}\label{section:Introduction}
\input{Text/1_Introduction}
\section{Methods}\label{section:Methods}
\input{Text/2_Methods}
\section{Results}\label{section:Results}
\input{Text/3_Results}
\section{Conclusion}\label{section:Conclusion}
\input{Text/4_Conclusion}
\section*{Acknowledgments}
\input{Text/5_Acknowledgements}
\FloatBarrier


\end{document}

%% file: Text/1_Introduction.tex
\lettrine{A}{s} engineering systems become increasingly complex, there is a need for design tools that can deliver effective solutions in pre-development \cite{IntroSPI2}. Systems such as hybrid battery packs \cite{HESS, HESSPLANE}, vehicle powertrain systems \cite{JvKampen}, low-cost tabletop MRI scanners \cite{MRI}, and compact infrastructure plants \cite{infra} benefit from improved design strategies to deliver desired solutions at lower societal cost.

Generative design leverages optimization and AI to explore design options, not as a one-step solution but as an extension of the design workflow, much like FEA (Finite Element Analysis) or CFD (Computational Fluid Dynamics). It automates the search through many configurations\cite{Borsboom}, helping balance competing requirements beyond what manual intuition can handle \cite{Andriesse}. However, when components interact strongly through shared space, this process needs an additional level of optimization \cite{Bello}.

Spatial Packaging of Interconnected Systems with Physical Interaction (SPI2) optimization frameworks address this by holistically evaluating the system as a whole \cite{ANovel}. By capturing mechanical, thermal, and electrical dependencies together, such frameworks can speed up the search for compact arrangements. It treats geometry, physical behavior, and routing as a single, coupled optimization problem rather than separate tasks \cite{Behzadi}. One unaddressed issue with current SPI2 frameworks is that components cannot yet fit into a custom boundary box \cite{Holistic}. 

Against this backdrop, this paper builds further on the framework presented in \cite{Steven} and implements a custom boundary box in which components can be fitted.

\begin{figure}[t]
\centering
\includegraphics[width=9cm]{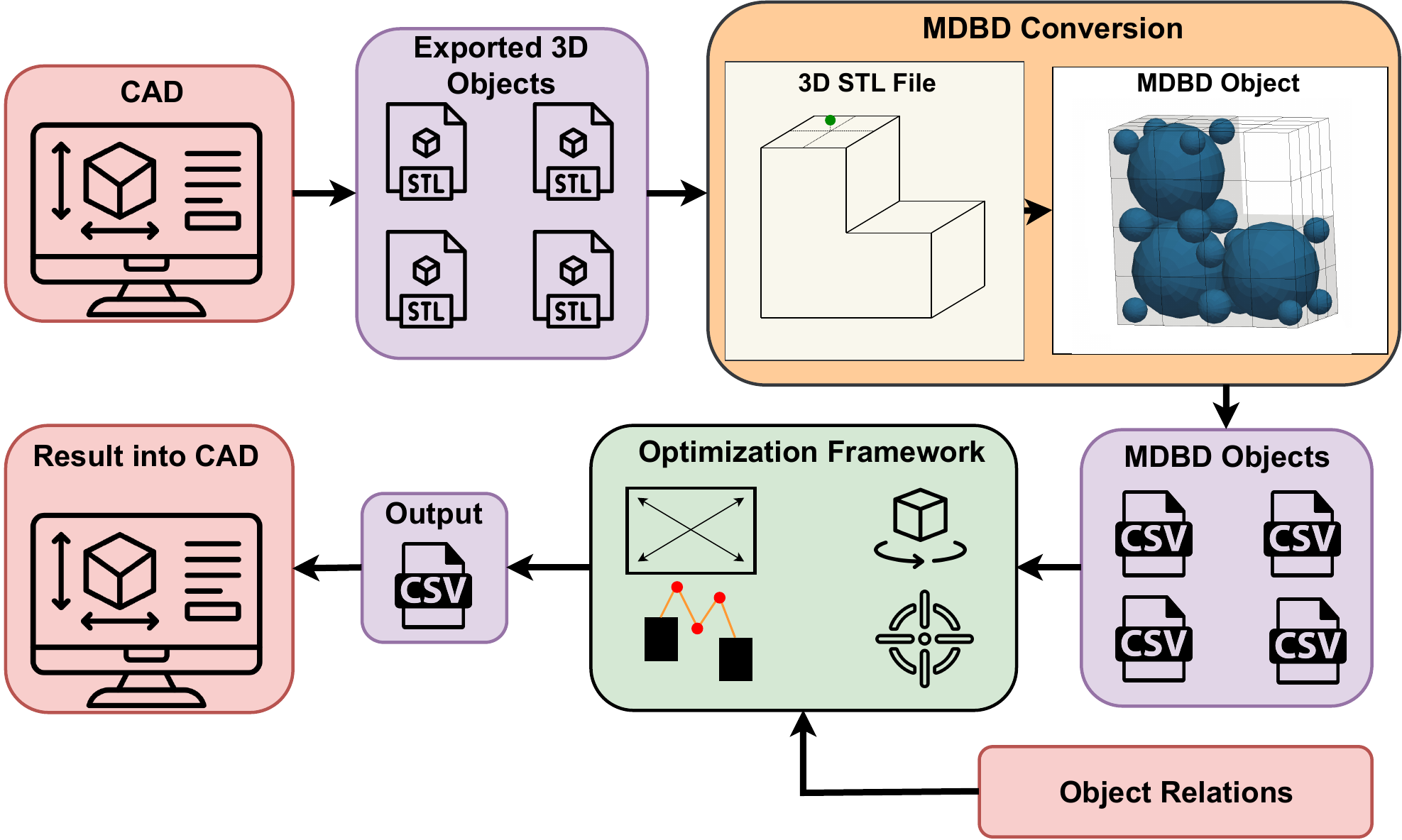}
\caption{Overview of the full framework from CAD drawing back to a CAD Drawing}
\label{fig:Pipeline}
\end{figure}

\subsubsection*{Related Literature}
This research relates to two research streams, SPI2 frameworks and methods for determining whether an object lies inside or outside a mesh.

The first research stream, SPI2. In this paper, we extend the framework introduced in \cite{Steven}, which focuses on spatial optimization for interconnected systems by combining routing and placement considerations. The framework leverages a geometric abstraction based on Maximal Disjoint Ball Decomposition (MDBD) \cite{Chen}. Its hybrid solving strategy, stochastic exploration followed by interior-point refinement, proved effective for non-convex, non-linear, constraint spatial problems.

SPI2 frameworks use an Axis-Aligned Bounding Box (AABB) \cite{Steven,Behzadi, Bello} to estimate the volume occupied by the components in the system. While this is a good measure for exploring how small a system can theoretically become, it is hardly the scenario in actual engineering applications. Consequently, current frameworks do not account for boundary conditions imposed by a custom-shaped design space. This limitation affects cases where the geometry must conform to product housings, vehicle envelopes, or architectural constraints.

In addition to the design boundary, we also include physics. The previously introduced framework \cite{Steven} has not included physics. Other SPI2 research has included physics-based thermal modeling \cite{Behzadi} and pressure loss in piping \cite{Bello}. The effects, such as center of gravity position and moment of inertia, are not yet implemented into SPI2 frameworks with the Maximal Disjoint Ball Decomposition representation.

The second research stream investigates when an object is inside or outside a mesh.

A well-researched and widely applied method to achieve this is the Ray Casting method \cite{Ray}. Which functions on the basis of casting a ray in a direction from a single point: if the number is odd, the point is inside the mesh; if the number is even, the point is outside the mesh \cite{Roth}. This poses the immediate problem that it would be hard to implement in a smooth, differentiable environment for optimization.  

The Signed Distance Function may overcome this limitation. This representation of geometry can represent the form of any three-dimensional shape by the magnitude of a point, where a region outside the object is positive and a region inside the object is negative, with zero indicating the boundary \cite{Takikawa}. This method of representing the design boundary might be very useful, as it yields a smooth, differentiable loss field \cite{Free-Surface}. This method has two disadvantages for our purpose. One is that the current optimization framework, which uses stochastic initialization with the interior-point optimizer IPOPT \cite{IPOPT} from CasADi \cite{CasADi}, does not natively support SDF tooling in Python. Second, traditionally, the more interior the point within the mesh, the lower the score; however, this is undesired in our application, as we do not want to favor “more” interior placement, since locating objects on the edges of the design space might be favorable for physics and routing. This is only a limitation when it is implemented as an objective function. So, why not use this as a constraint? The reason is that in “tight” design spaces, all stochastically sampled initial positions might fall outside the design space, leaving no suitable starting position. There are many good solutions. To prevent this, we want to enable initialization outside the design space and let the interior-point optimization find an optimal position within the design boundary from there.

With these reasons in mind, we set out to develop a new method that leverages the inherent properties of MDBD to create a smooth, differentiable method for determining whether objects are within or outside a custom-shaped design space.

In conclusion, to the best of the author’s knowledge, there are no SPI2 optimization frameworks that account for a custom non-convex design space.  

\subsubsection*{Statement of Contributions}
This paper presents an optimization methodology to automatically position objects and routing, considering the physical interaction of center of mass and moment of inertia, within a specified design boundary. The work introduces a novel method to determine if a MDBD object is located within the design boundary or outside the design boundary. In addition this framework is placed in a workflow where the components can be setup in CAD software, and the resultant assembly design can be imported to CAD software to evaluate the results. The full framework is shown in Fig.~\ref{fig:Pipeline}.

\subsubsection*{Organization}
The structure of the paper is as follows. Section~\ref{section:Methods} defines the optimization framework. Section~\ref{section:Results} applies the proposed framework to a demonstration problem. Section~\ref{section:Conclusion} concludes the work and suggests directions for future research.

%% file: Text/2_Methods.tex
%
\subsection{MDBD Model}\label{sec:MDBD}

\begin{figure}[t]
\centering
\includegraphics[width=9cm]{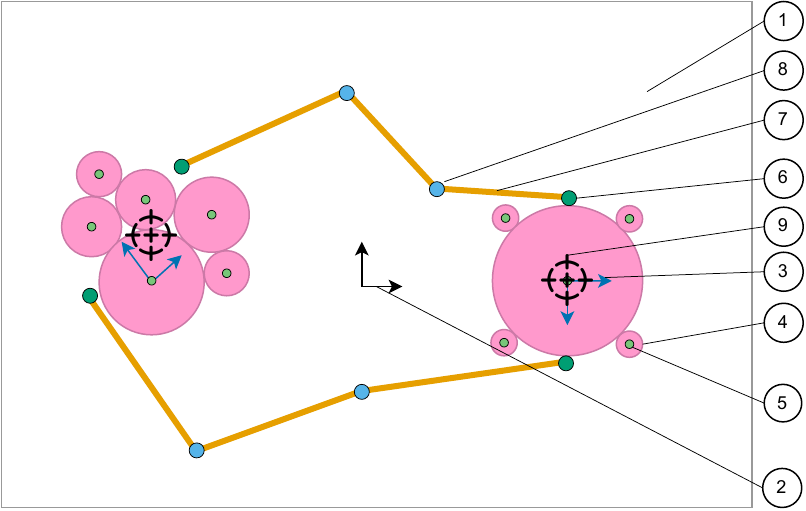}
\caption{Depiction of the model with various components: \(n_\mathrm{d}\)-dimensional workspace \(\mathbb{W}\) (1, White), Cartesian frame \(\mathbb{F}_\mathbb{W}\) (2, Black), Object Frame \(\mathbb{F}_{A_i}\) (3, Dark Blue), Spheres \(b_{i,\mu}\) (4, Light Purple), center \(\mathbf{p}_{\mathrm{b}_{i,\mu}}\) (5, Light Green), ports \(\varphi_{i,\ell}\) (6, Dark Green), Routing segments (7, Orange), control points \(\mathbf{c}_{L,k}\) (8, Light Blue), and the center of gravity \(\mathbf{p}_{\mathrm{CoG},{i}}\) (9, Black). An adjustment from \cite{Steven}.}
\label{fig:DesignSpace}
\end{figure}

For the geometric representation of objects, we use the Maximal Disjoint Ball Decomposition as introduced by \cite{Chen}. With this methods we can create a smooth and differentiable optimization problem for placement, routing and physics. The setup of the model is equal to \cite{Steven}, with an adjustment to include the center of gravity. Each object $A_i$ is represented as a collection of non-overlapping spheres that collectively approximate its geometry. Formally, $A_i$ consists of the set
\begin{equation}
A_i = \{\, b_{i,1}, b_{i,2}, \dots, b_{i,\,n_{\mathrm{b},i}} \,\},
\end{equation}
where every sphere $b_{i,\mu}$ is characterized by its center position $\mathbf{p}_{\mathrm{b},i,\mu}$ and corresponding radius $r_{\mathrm{b},i,\mu}$.
The spheres within a single object are constrained to remain pairwise disjoint, such that $b_{i,\mu} \cap b_{i,\nu} = \emptyset$ for all $\mu \neq \nu$ \cite{Chen2}. The parameter $n_{\mathrm{b},i}$ denotes the number of spheres used in the decomposition, thereby defining the geometric resolution of object $i$. This representation fully specifies the object through its set of sphere centers and radii, providing a compact and differentiable geometric description suitable for optimization \cite{Behzadi}.

We consider an $n_\mathrm{d}$-dimensional workspace $\mathbb{W}$ with a global Cartesian reference frame $\mathbb{F}_\mathbb{W}$, as illustrated in Fig.~\ref{fig:DesignSpace}. The environment contains $n_\mathrm{obj}$ rigid bodies, where each object $A_i$ is associated with a local coordinate frame $\mathbb{F}_{A_i}$ and is modeled as a set of $n_{\mathrm{b},i}$ disjoint spheres $b_{i,\mu}$. In addition, objects may include interface ports $\varphi_{i,\ell}$ for connectivity, where $\ell = 1,\dots,n_{\varphi,i}$ and $n_{\varphi,i}$ denotes the number of ports on object $i$. Routing paths are constructed between ports and intermediate control points $\mathbf{c}_{L,k}$, which define linear connections between components. Each route $L$ consists of $K_L$ straight segments that connect $K_L + 1$ nodes \([\mathbf{q}_{L,0},\mathbf{q}_{L,1},\dots,\mathbf{q}_{L,K_L}]\), where the endpoints \(\mathbf{q}_{L,0}\) and \(\mathbf{q}_{L,K_L}\) are ports and the intermediate nodes \(\mathbf{q}_{L,k}(k=1,\dots,K_L - 1)\) are control points \(\mathbf{c}_{L,k}\). The number of segments is \(K_L = n_{\mathrm{cp},L}+1\), and segments \(m\) connect nodes \(\mathbf{q}_{L,m}\) and \(\mathbf{q}_{L,m+1}\) for \(m=0,\dots,K_L-1\). Where \(n_{\mathrm{cp},L}\) is the total number of control points across a single route \(L\). The position of the center of gravity is denoted by \(\mathbf{p}_{\mathrm{CoG},{i}}\), the mass of object \(A_i\) is \(m_{\mathrm{CoG},{i}}\).

For the remainder of this study, we set $n_\mathrm{d}=3$. The position of each object is represented by the parameter vector
\begin{equation}
\mathbf{x}_{A_i} = [x_{\theta,i},\, x_{\alpha,i},\, x_{\beta,i},\, x_{x,i},\, x_{y,i},\, x_{z,i}]^\top \in \mathbb{R}^6,
\end{equation}
where $(x_{\theta,i},x_{\alpha,i},x_{\beta,i})$ correspond to the rotational degrees of freedom (roll, pitch, yaw) and $(x_{x,i},x_{y,i},x_{z,i})$ to the translational components in the workspace frame.
Control points contain only positional information and are represented as
\begin{equation}
\mathbf{c}_{L,k} = [c_{x,L,k},\, c_{y,L,k},\, c_{z,L,k}]^\top \in \mathbb{R}^3.
\end{equation}

The rigid transformation from an object's local frame $\mathbb{F}_{A_i}$ to the workspace frame $\mathbb{F}_\mathbb{W}$ is given by
\begin{equation}
\mathbf{p}^{\mathbb{W}} = \mathbf{R}_i\,\mathbf{p}^{A_i} + \mathbf{t}_i,
\quad
\mathbf{R}_i \in \mathbb{R}^{3\times 3}, \ \ \mathbf{t}_i \in \mathbb{R}^{3},
\end{equation}
where $\mathbf{p}^{A_i}$ may represent the position of a sphere center $\mathbf{p}_{\mathrm{b}_{i,\mu}}$ or a routing node $\mathbf{q}_{L,K_L}$.
Rotations are expressed using Roll--Pitch--Yaw (RPY) angles~\cite{Multibod}, with
\begin{equation}
\mathbf{R}_i = \mathbf{R}_z(x_{\theta,i})\,\mathbf{R}_y(x_{\alpha,i})\,\mathbf{R}_x(x_{\beta,i}).
\end{equation}
The full vector of design variables is then defined as
\begin{equation}
\mathbf{x} = [\,\mathbf{x}_1^\top, \dots, \mathbf{x}_{n_\mathrm{obj}}^\top,\, \mathbf{c}_{1,1}^\top, \dots, \mathbf{c}_{L,k}^\top\,]^\top.
\end{equation}

\subsection{Constraints}
Each sphere is not allowed to intersect with another sphere or any routing segment. 
Representing geometry with spheres offers an advantage, pairwise interference checks 
are computationally lightweight, even though the overall placement–routing problem remains non-convex.
Each sphere \(b_{i,\mu}\) has a center 
\(\mathbf{p}^{\mathbb{W}}_{\mathrm{b}_{i,\mu}}\in\mathbb{R}^{3}\) 
and radius \(r_{\mathrm{b}_{i,\mu}}\).

For every object pair \((i,j)\) with \(i<j\), and for all sphere indices 
\(\mu=1,\ldots,n_{\mathrm{b},i}\) and \(\nu=1,\ldots,n_{\mathrm{b},j}\), 
the signed clearance between two spheres is defined as
\begin{equation}
  d^{\,\mathrm{obj\mbox{-}obj}}_{i,\mu,j,\nu}
  \;=\;
  \left\|\,\mathbf{p}^{\mathbb{W}}_{\mathrm{b}_{i,\mu}}
           -\mathbf{p}^{\mathbb{W}}_{\mathrm{b}_{j,\nu}}\,\right\|_2
  \;-\;
  \bigl(r_{\mathrm{b}_{i,\mu}}+r_{\mathrm{b}_{j,\nu}}\bigr),
  \label{eq:clearance_objobj}
\end{equation}
similar to \cite{Steven,Behzadi}.
Non-overlap requires
\begin{equation}
  d^{\,\mathrm{obj\mbox{-}obj}}_{i,\mu,j,\nu}\;\ge\;0,
  \qquad
  \forall\, i<j,\;\mu=1,\dots,n_{\mathrm{b},i},\;\nu=1,\dots,n_{\mathrm{b},j}.
  \label{eq:nonoverlap_objobj}
\end{equation}

where the total number of sphere-pair constraints is
\begin{equation}
  N_{\mathrm{pairs}}
  \;=\;
  \sum_{1\le i<j\le n_\mathrm{obj}} n_{\mathrm{b},i}\,n_{\mathrm{b},j}.
\end{equation}

All pairwise clearances are collected in the constraint vector
\begin{equation}
  \mathbf{g}^{\mathrm{obj\mbox{-}obj}}
  \;=\;
  -\bigl[\,d^{\,\mathrm{obj\mbox{-}obj}}_{i,\mu,j,\nu}\,\bigr]_{(i,j,\mu,\nu)}
  \;\in\;\mathbb{R}^{N_{\mathrm{pairs}}},
\end{equation}
such that feasibility corresponds to
\begin{equation}
    \mathbf{g}^{\mathrm{obj\mbox{-}obj}}(\mathbf{x}) \;\leq\; 0.
\end{equation}

In addition, routing segments must not intersect with objects or with each other. 
These constraints follow the formulations in \cite{Steven, Chen2, Behzadi} 
and are stated in their negative-null (IPOPT-compatible) form:
\begin{equation}
    \mathbf{g}^{\mathrm{route\text{-}obj}}(\mathbf{x})\;\leq\; 0,
\end{equation}
\begin{equation}
    \mathbf{g}^{\mathrm{route\text{-}route}}(\mathbf{x})\;\leq\; 0.
\end{equation}

Taken together, these non-intersection constraints ensure that objects and routing segments remain mutually separated in space. Using the same geometric sphere-based representations and smooth distance-based measures, we can also enforce the complementary requirement that an object remains \emph{inside} another, enclosing geometry. This idea naturally leads to the formulation of the design boundary presented next.

\subsection{Design Boundary}
\begin{figure}
    \centering
    \includegraphics[width=7cm]{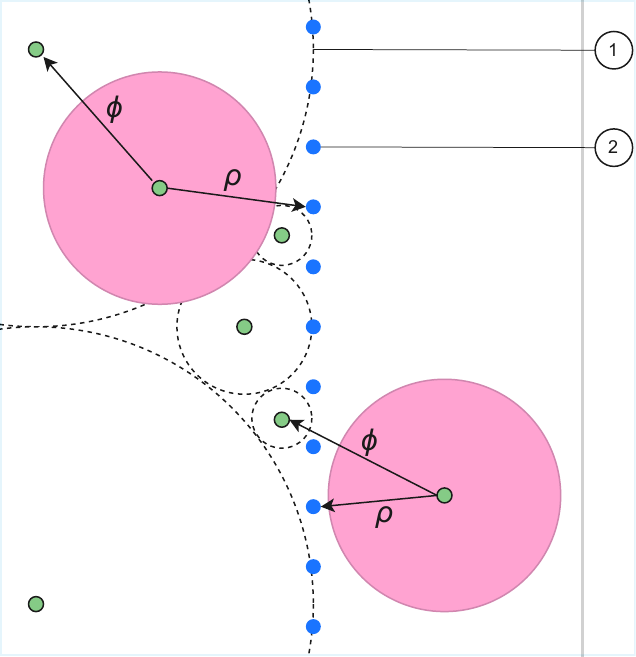}
    \caption{Example of a design boundary section, where the boundary spheres $b_{\mathrm{B},\mu}$ (1, dotted outline) and boundary points $\mathbf{p}_{\mathrm{Box},\mu}$ (2, blue) are shown. In addition, two object spheres are shown, one within the design boundary and one outside the design boundary, illustrating the relations of $\phi$ and $\rho$.}
    \label{fig:DesignBoundary}
\end{figure}

To ensure that all objects remain within the predefined design boundary, a smooth boundary objective is formulated based on soft minimum and smooth hinge operators.

The design boundary can represent any enclosing component that constrains the placement of other components, such as the engine bay of a vehicle or the fuselage of an aircraft. This boundary is modeled as a watertight 3D geometry, which is converted into a Maximal Disjoint Ball Decomposition (MDBD) representation. In addition to boundary spheres, a set of mesh points is sampled from the surface of the 3D model to describe the boundary envelope more precisely. A schematic overview of this setup is shown in Fig.~\ref{fig:DesignBoundary}.

We define this boundary representation as
\begin{equation}
    A_{\mathrm{B}} = \{\, b_{\mathrm{B},1},\dots, b_{\mathrm{B},n_{\mathrm{B}}},\, 
    \mathbf{p}_{\mathrm{Box},1},\dots, \mathbf{p}_{\mathrm{Box},n_{\mathrm{Box}}} \,\},
\end{equation}
where $b_{\mathrm{B},\mu}$ denotes a boundary sphere with center $\mathbf{p}_{\mathrm{bB},\mu}$ and radius $r_{\mathrm{bB},\mu}$, and 
$\mathbf{p}_{\mathrm{Box},\mu}$ represents a point on the boundary mesh. 
Using these definitions, the spatial relationship between an object $A_i$ and the design boundary $A_{\mathrm{B}}$ can be expressed as
\begin{equation}
    \phi_{\mu,i,\nu} = \|\mathbf{p}_{\mathrm{bB},\mu} - \mathbf{p}_{\mathrm{b},i,\nu}\|_2 - r_{\mathrm{bB},\mu},
\end{equation}
which becomes negative when the center of an object sphere $\mathbf{p}_{\mathrm{b},i,\nu}$ lies within a boundary sphere $b_{\mathrm{B},\mu}$.

Additionally, the object sphere $b_{i,\nu}$ must not intersect the boundary surface defined by the mesh points $\mathbf{p}_{\mathrm{Box},\mu}$,
\begin{equation}
    \rho_{\mu,i,\nu} = \|\mathbf{p}_{\mathrm{Box},\mu} - \mathbf{p}_{\mathrm{b},i,\nu}\|_2 - r_{\mathrm{b},i,\nu}.
\end{equation}
Here, a positive value of $\rho_{\mu,i,\nu}$ indicates that the object sphere does not intersect the boundary surface.

These two relations define the conditions required for an object to be fully enclosed within the design boundary. Specifically, all $\phi_{\mu,i,\nu}$ must be negative, ensuring that every object sphere lies within the boundary, while all $\rho_{\mu,i,\nu}$ must be positive, ensuring no intersection occurs. This can be summarized as
\begin{equation}
    \phi_{\mathrm{max}} = \max_{\forall\mu,i,\nu}\big(\phi_{\mu,i,\nu}\big),
    \quad
    \rho_{\mathrm{min}} = \min_{\forall\mu,i,\nu}\big(\rho_{\mu,i,\nu}\big),
\end{equation}
and the inclusion condition can be expressed as
\begin{equation}
    \phi_{\mathrm{max}} \leq 0 \quad \text{and} \quad \rho_{\mathrm{min}} \geq 0.
\end{equation}

Since the max/min operators are non-smooth, we introduce smooth log-sum-exp operators to obtain differentiable approximations \cite{LogSumExp}. 
Let \(v = \{v_k\}_{k=1}^{n_v}\) denote a finite set of scalar values \(v_k \in \mathbb{R}\), and let \(\alpha > 0\) be a smoothing parameter. 
The soft maximum and soft minimum of the set \(v\) are defined respectively as
\begin{equation}
\begin{split}
    M^+(v,\alpha) &= \tfrac{1}{\alpha}\log\!\Big(\sum_{k=1}^{n_v} e^{\alpha v_k}\Big),\\[2pt]
    M^-(v,\alpha) &= -\tfrac{1}{\alpha}\log\!\Big(\sum_{k=1}^{n_v} e^{-\alpha v_k}\Big),
\end{split}
\end{equation}
where the parameter \(\alpha\) controls the degree of smoothness: 
larger values of \(\alpha\) yield a closer approximation to the exact maximum and minimum operators.
 Accordingly, for each object sphere $(i,\nu)$ we define smooth envelope functions
\begin{equation}
\begin{split}
    \tilde{\phi}_{i,\nu} 
    &= M^+\big(\{\phi_{\mu,i,\nu}\}_{\mu=1}^{n_\mathrm{B}},\,\alpha_{\mathrm{U}}\big),\\[3pt]
    \tilde{\rho}_{i,\nu} 
    &= M^-\big(\{\rho_{\mu,i,\nu}\}_{\mu=1}^{n_\mathrm{Box}},\,\alpha_{\mathrm{P}}\big),
\end{split}
\end{equation}
where $\alpha_{\mathrm{U}}$ and $\alpha_{\mathrm{P}}$ denote the smoothness parameters for the union and point envelopes, respectively.

To ensure numerical stability and differentiability near constraint boundaries, both terms are passed through a smooth hinge function defined as
\begin{equation}
    H(v,\beta) = \tfrac{1}{\beta}\log\!\big(1 + e^{\beta v}\big),
\end{equation}
where $\beta$ defines the sharpness of the transition \cite{Softplus}. This results in the per-sphere penalty terms
\begin{equation}
\begin{split}
    \Psi^{\text{union}}_{i,\nu} &= H\!\big(\tilde{\phi}_{i,\nu} + \delta_u,\,\beta\big),\\[2pt]
    \Psi^{\text{points}}_{i,\nu} &= H\!\big(-\tilde{\rho}_{i,\nu} - \delta_p,\,\beta\big),
\end{split}
\end{equation}
where $\delta_u$ and $\delta_p$ provide small safety margins for inner and outer violations, respectively.

Finally, the overall smooth boundary objective is computed as a weighted sum:
\begin{equation}
    f_{\mathrm{B}}(\mathbf{x})
    = \sum_{i}\sum_{\nu}
      \Big(
        w_{\mathrm{union}}\,\Psi^{\text{union}}_{i,\nu}
        + w_{\mathrm{points}}\,\Psi^{\text{points}}_{i,\nu}
      \Big),
\end{equation}
where $w_{\mathrm{union}}$ and $w_{\mathrm{points}}$ define the relative weight of the two penalties. The objective $f_\mathrm{B}$ depends directly on the design vector $\mathbf{x}$ through the translated object spheres, as described in Section~\ref{sec:MDBD}. By including $f_\mathrm{B}(\mathbf{x})$ in the optimization objective, we encourage solutions that remain within the design boundary.

\subsubsection*{An Omission}
Now, there is an edge case in which an object sphere is within the design boundary but the center point \(\mathbf{p}_{\mathrm{b},i,\nu}\) is outside one of the design boundary spheres. An additional rule could be added to fix this, however, when the resolution of the spheres in the design boundary is high enough the framework functions as desired.

\subsection{Physics}
\subsubsection*{Center of Gravity}
For this SPI2 setup we consider the center of gravity and the moment of inertia.
The global center of gravity (CoG) of the assembly is computed as the mass-weighted average of the individual object centers of gravity after transformation to world coordinates. 
Let $n_\mathrm{obj}$ denote the number of objects, and let each object $i$ have a mass $m_{\mathrm{CoG},{i}}$ and a transformed position of its local CoG $\mathbf{p}_{\mathrm{CoG},{i}} \in \mathbb{R}^3$. 
The global CoG, denoted as $\mathbf{p}_\mathrm{G}$, is obtained as

\begin{equation}
\mathbf{p}_\mathrm{G} = 
\frac{\sum_{i=1}^{n_\mathrm{obj}} m_{\mathrm{CoG},{i}} \, \mathbf{p}_{\mathrm{CoG},{i}}}
     {\sum_{i=1}^{n_\mathrm{obj}} m_{\mathrm{CoG},{i}}}.
\label{eq:CoG}
\end{equation}

Equation~\eqref{eq:CoG} defines a continuous and differentiable mapping suitable for symbolic evaluation in CasADi. 
The numerator represents the first moment of mass, while the denominator normalizes by the total mass.

During optimization, the global CoG can be guided toward a desired reference point $\mathbf{p}_\mathrm{G}^{*}$ (specified design target). The CoG deviation objective is formulated as
\begin{equation}
f_{\mathrm{CoG}}(\mathbf{x}) =
\left\| 
\mathbf{p}_\mathrm{G} - \mathbf{p}_\mathrm{G}^{*}
\right\|_2^2,
\label{eq:CoG_obj}
\end{equation}

which penalizes quadratic deviations between the actual and desired CoG positions. 
This term is incorporated into the total objective function, allowing the nonlinear optimizer (IPOPT) to adjust object positions and orientations such that the assembly’s overall CoG aligns with the target location.

\subsubsection*{Moment of Inertia About the Global CoG}

In the SPI2 setup, the moment of inertia is always evaluated about the global center of gravity $\mathbf{p}_\mathrm{G}$ of the assembly, as defined in~\eqref{eq:CoG}. Each object $A_i$ is modeled as a rigid body with mass $m_{\mathrm{CoG},i}$, local CoG position $\mathbf{p}_{\mathrm{CoG},i} \in \mathbb{R}^3$ expressed in the workspace frame, orientation $\mathbf{R}_i \in \mathbb{R}^{3\times 3}$ mapping the object frame $\mathbb{F}_{A_i}$ to the workspace frame $\mathbb{F}_\mathbb{W}$, and inertia tensor $\mathbf{I}^{\text{body}}_i$ expressed about the local CoG. Let $\mathbf{I}_3$ denote the $3\times 3$ identity matrix. The calculation for the moment of inertia is a variation on the methods presented in \cite{Multibod}.

The inertia of object $A_i$ about its own CoG but expressed in the workspace frame is
\begin{equation}
    \mathbf{I}^{\mathbb{W},\mathrm{CoG}}_i 
    = \mathbf{R}_i\,\mathbf{I}^{\text{body}}_i\,\mathbf{R}_i^\top.
    \label{eq:inertia_rotated_cas}
\end{equation}

The vector from the global CoG to the CoG of object $A_i$ is
\begin{equation}
    \mathbf{d}_i = \mathbf{p}_{\mathrm{CoG},i} - \mathbf{p}_\mathrm{G}.
    \label{eq:inertia_di_vector}
\end{equation}
Applying the parallel-axis theorem, the inertia of object $A_i$ about $\mathbf{p}_\mathrm{G}$ becomes
\begin{equation}
    \mathbf{I}^{\mathbb{W},\mathbf{p}_\mathrm{G}}_i 
    = \mathbf{I}^{\mathbb{W},\mathrm{CoG}}_i 
      + m_{\mathrm{CoG},i}
        \big( \|\mathbf{d}_i\|_2^2 \mathbf{I}_3 - \mathbf{d}_i \mathbf{d}_i^\top \big).
    \label{eq:inertia_parallel_cas}
\end{equation}

Summing the contributions of all objects yields the total inertia about the global CoG:
\begin{equation}
\begin{split}
    \mathbf{I}_{\text{tot}}(\mathbf{p}_\mathrm{G})
    &= \sum_{i=1}^{n_\mathrm{obj}}
       \mathbf{I}^{\mathbb{W},\mathbf{p}_\mathrm{G}}_i \\[2pt]
    &= \sum_{i=1}^{n_\mathrm{obj}}
       \big[
           \mathbf{R}_i\,\mathbf{I}^{\text{body}}_i\,\mathbf{R}_i^\top
           + m_{\mathrm{CoG},i}
             \big( \|\mathbf{d}_i\|_2^2 \mathbf{I}_3 - \mathbf{d}_i \mathbf{d}_i^\top \big)
       \big].
\end{split}
\label{eq:inertia_about_global_cog}
\end{equation}
All operations in~\eqref{eq:inertia_rotated_cas}–\eqref{eq:inertia_about_global_cog} are smooth and suitable for automated differentiation in CasADi.

When $\mathbf{I}_{\text{tot}}(\mathbf{p}_\mathrm{G})$ is written in the workspace frame, its diagonal entries correspond to the principal moments about the world axes:
\begin{equation}
\begin{aligned}
    I_{xx} &= \big(\mathbf{I}_{\text{tot}}(\mathbf{p}_\mathrm{G})\big)_{11}, \\[2pt]
    I_{yy} &= \big(\mathbf{I}_{\text{tot}}(\mathbf{p}_\mathrm{G})\big)_{22}, \\[2pt]
    I_{zz} &= \big(\mathbf{I}_{\text{tot}}(\mathbf{p}_\mathrm{G})\big)_{33}.
\end{aligned}
\label{eq:inertia_diagonal_cas}
\end{equation}

A scalar inertia objective is defined as
\begin{equation}
    f_{\mathrm{I}}(\mathbf{x})
    = w_{x} I_{xx} + w_{y} I_{yy} + w_{z} I_{zz},
    \label{eq:inertia_obj_cas}
\end{equation}
where $w_x$, $w_y$, and $w_z$ are user-defined weights. This term is added to the total objective to favor configurations with desirable inertia properties around the global CoG.

\subsection{Routing Length}
The determination of the routing length is equal to \cite{Steven}. The total routing length is calculated by summing all routing segments as
\begin{align}
f_\mathrm{r}(\mathbf{x}) &=
\sum_{L}\sum_{m=0}^{K_L-1} 
   \bigl\| \mathbf{q}_{L,m+1} - \mathbf{q}_{L,m} \bigr\|_2^2,
\label{eq:fr}
\end{align}
Where the distances are squared to prevent undetermined situations.

\subsection{Optimization Problem}
Now using the previously defined objective functions, we can combine this into one optimization problem:
\begin{prob}\label{prob:1}
Optimization of the SPI2 problem with a design boundary and physics.
\begin{equation*}
\begin{aligned}
& \min && f(\mathbf{x}) = \omega_\mathrm{r} f_\mathrm{r}(\mathbf{x})+\omega_\mathrm{B} f_\mathrm{B}(\mathbf{x})+ \omega_\mathrm{CoG} f_\mathrm{CoG}(\mathbf{x})+\omega_\mathrm{I} f_\mathrm{I}(\mathbf{x})\\
& \text{s.t.} && \mathbf{g}^{\mathrm{obj\mbox{-}obj}}(\mathbf{x}) \leq 0,\\
& && \mathbf{g}^{\mathrm{route\text{-}obj}}(\mathbf{x})  \leq 0, \\
& && \mathbf{g}^{\mathrm{route\mbox{-}route}}(\mathbf{x})  \leq 0.
\end{aligned}
\end{equation*}
\end{prob}

\begin{figure}[t]
\centering
\includegraphics[width=7.5cm]{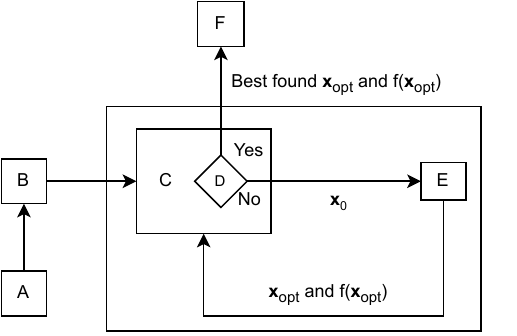}
\caption{Overview of the nested optimization framework. Steps A--B construct the MDBD model and initialize object and routing variables. Step C selects an initialization method. Step D checks the exit criterion. Step E performs IPOPT-based optimization. Step F outputs the final placement and routing results.\cite{Steven}}
\label{fig:NestedApproach}
\end{figure}

\subsection{Nested Optimization Framework}

The placement and routing problem addressed in this work is constrained, nonlinear, and non-convex. As a result, gradient-based nonlinear programming (NLP) methods are required. We use the interior-point solver IPOPT from CasADi~\cite{CasADi, IPOPT}, which offers efficient constraint handling, Lagrange multiplier information, and second-order convergence properties suitable for large sets of spatial and physical constraints.

Because the problem is non-convex, a single NLP solve may converge to an undesirable local optimum. To reduce this risk, a \emph{nested} strategy is adopted. In this approach, IPOPT is repeatedly initialized from different starting points, with each iteration producing a new candidate solution. Each initial guess~$\mathbf{x}_0$ is refined by IPOPT into a locally optimal solution~$\mathbf{x}_\mathrm{opt}$, and the resulting objective value is stored. After a fixed number of initializations, the best-performing solution is selected as the final configuration.

In this work, all initial guesses are generated using a \textit{random initialization} strategy. For each iteration, the design vector~$\mathbf{x}_0$ is sampled uniformly within predefined bounds for object positions and routing control points. This produces a diverse set of starting configurations, enabling broad exploration of the solution space before local refinement by IPOPT.

Because the random samples cover the domain without structural bias, the nested procedure can identify high-quality placements by evaluating many locally optimized candidates. This approach avoids premature convergence, and achieves reliable performance across varying problem instances \cite{Steven}.

As shown in Fig.~\ref{fig:NestedApproach}, each iteration of the nested approach proceeds as follows:
(A-B) construct the MDBD objects and routing representation,  
(C-D) generate a random initial design vector~$\mathbf{x}_0$,  
(E) solve the corresponding NLP using IPOPT, and  
(F) store the optimized design~$\mathbf{x}_\mathrm{opt}$ and its objective value.  

This loop continues until a predefined number of random initializations has been evaluated. The best candidate across all iterations is then selected, yielding the final arrangement of object placements and routing paths.

\subsection{CAD to CAD}
One important aspect of a design tool is its applicability. In this framework, we create a loop in which a design is started in Computer-Aided Design (CAD), and the framework provides an output that results in a CAD assembly. The CAD software FreeCAD \cite{FreeCAD} is used. Geometric transformations are performed by creating a homogeneous reference frame for each object, which is maintained after each conversion. The optimization provides a final result, $\mathbf{x}_\mathrm{opt}$, containing the design vectors for the objects and control points. These translate one-to-one back to their respective parts in CAD. In the results section, we provide an example in which the full SPI2 framework is used and a few simple components are combined by the optimization framework into an assembly. One drawback is that these transformations are done manually at this point.

%% file: Text/3_Results.tex
\begin{figure}
    \centering
    \includegraphics[width=9cm]{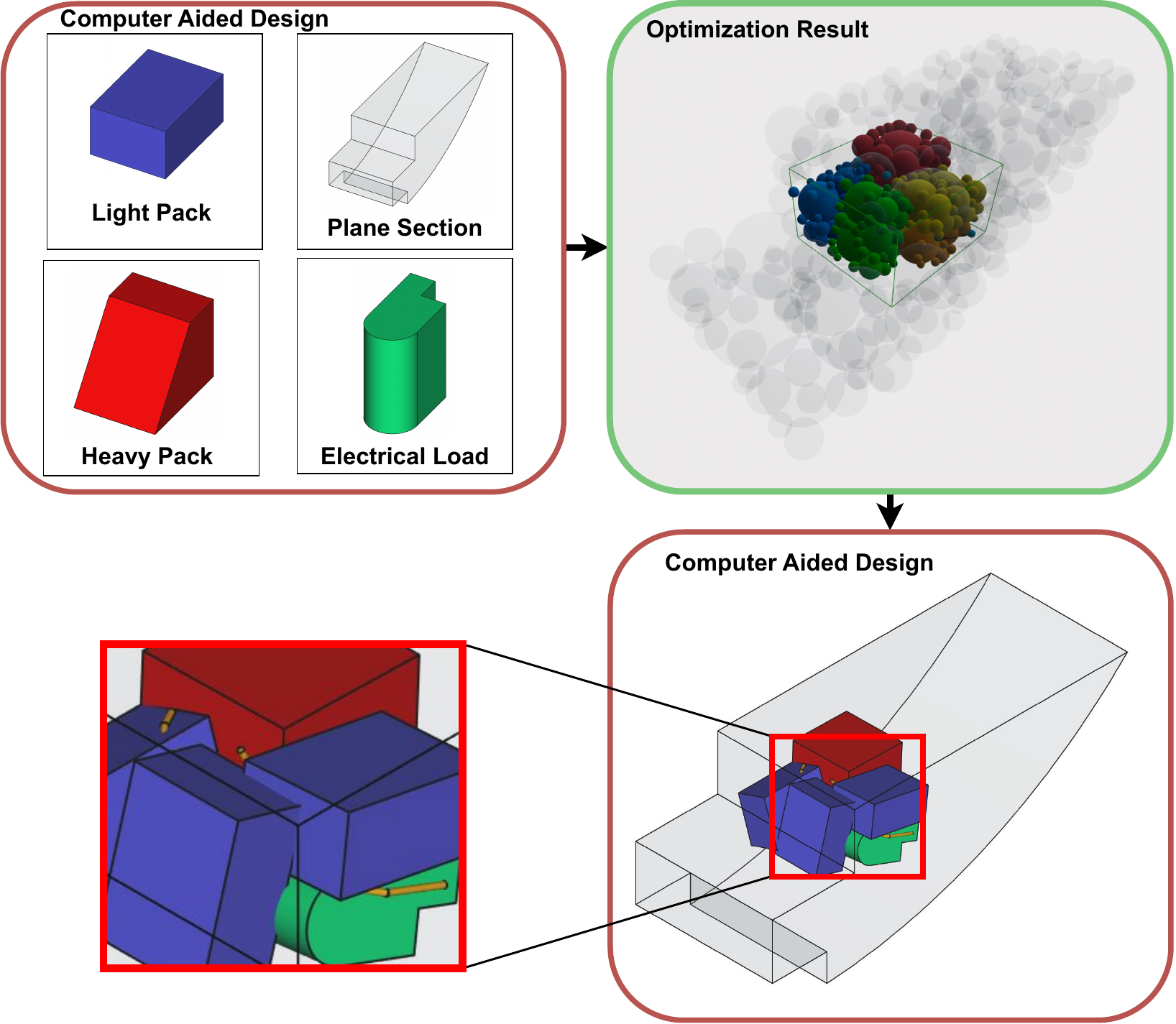}
    \caption{Overview of the placement–routing–physics outcome generated by the proposed optimization framework. Top left: initial CAD parts. Top right: locally optimal MDBD arrangement. Bottom right: CAD-based geometric reconstruction of the solution. Bottom left: close-up view of the components. Interconnections are shown in orange.}
    \label{fig:HESS_Plane}
\end{figure}

In this section, we evaluate the framework using a fictional auxiliary unit located in the rear section of an aircraft. Three parts and a tail section were created in FreeCAD \cite{FreeCAD} and converted into MDBD objects: three light packs, one heavy pack, and an electrical load. Each object is represented by 40 MDBD spheres and sequentially connected using cylindrical routing segments. All objective weights were set to one, and the problem was evaluated on routing length, design boundary utilization, center of gravity, and moment of inertia.

Experiments were performed on a 64-bit Windows laptop with an Intel\textsuperscript{\textregistered} Core\textsuperscript{TM} i7-7700HQ @ 2.80 GHz, 8 GB RAM, a 238 GB Samsung SSD, and an NVIDIA Quadro M1200 (4 GB) alongside Intel\textsuperscript{\textregistered} HD Graphics 630. The framework was implemented in Python~3.11.9 \cite{Python}. The problem was randomly initialized and solved 75 times using an interior-point method; the best solution is shown in Fig.~\ref{fig:HESS_Plane}. The total runtime was approximately 5 hours and 15 minutes.

The normalized objective values of the selected result are: routing length $0.61\,[-]$, bounding box $0.46\,[-]$, center of gravity $0.11\,[-]$, and moment of inertia $3.50\,[-]$, giving a total of $4.68\,[-]$. From the constraint outputs, the minimum sphere-to-sphere distance is $-1.19 \times 10^{-8}\,[-]$, indicating negligible interference and within numerical tolerance. This confirms that neither the spheres nor the routing segments exhibit significant overlap that would invalidate the solution.

Figure~\ref{fig:HESS_Plane} shows the optimized MDBD placement and its CAD reconstruction. The bottom-left close-up highlights several observations. First, all objects are successfully placed within the non-convex design boundary. Second, slight collisions are visible between CAD parts, caused by the relatively low spherical resolution used in the MDBD approximations. Finally, most routing segments run between the objects as expected, though some segments intersect CAD surfaces due to the simplified geometric abstraction used during optimization.

%% file: Text/4_Conclusion.tex
In this paper, we introduced a novel method for placing objects within a custom non-convex design space in the context of Spatial Packaging of Interconnected Systems with Physical Interaction (SPI2). In addition, the paper presents a first step toward integrating the framework into a full Computer-Aided Design (CAD) workflow, enabling multiple components to be optimally arranged and then reconstructed as a CAD assembly.

The proposed approach extends existing SPI2 frameworks in several ways. First, it introduces a smooth, differentiable boundary-checking method based on the geometric properties of the Maximal Disjoint Ball Decomposition (MDBD), enabling gradient-based optimization to handle custom-shaped design spaces. Second, the framework incorporates physical metrics such as center of gravity and moment of inertia. Third, by supporting both MDBD abstraction and CAD reconstruction, the method establishes a continuous pipeline from CAD models to optimization and back to CAD, providing a first step in closing the loop needed for practical engineering workflows.

The demonstration problem illustrates that the method can successfully position interconnected components inside a non-convex boundary while simultaneously optimizing routing length, spatial utilization, and physical behavior. The results confirm that the proposed formulation produces feasible assemblies, and can handle non-convex geometries.

The following extensions to this work are possible. First, the process from CAD to CAD can be fully automated to speed up result translation. Second, due to the problem's non-linear and non-convex nature, optimal results cannot be guaranteed; however, the ultimate goal is to improve design quality and engineer efficiency. A test can be conducted in which simple designs are created by a large number of engineering students and evaluated against the framework to obtain a true measure of the optimization framework's quality and speed.

%% file: Text/5_Acknowledgements.tex
This report has been partially prepared with the assistance of generative AI tools (OpenAI ChatGPT). These tools were used to improve the structure, clarity, and consistency of the writing, as well as to format the LaTeX document. All ideas, analysis, and technical content originate from the authors and the final text has been reviewed and verified for accuracy and originality.